\documentstyle[aps,prb,twocolumn,epsf,floats]{revtex}

\begin{document}

\draft
\wideabs{

\title{Commensurate-Incommensurate transition in the melting process \\
of the orbital ordering in Pr$_{0.5}$Ca$_{0.5}$MnO$_{3}$: neutron
diffraction study}

\author{R. Kajimoto}

\address{Department of Physics, Faculty of Science, Ochanomizu
University, Bunkyo-ku, Tokyo 112-8610, Japan}

\author{H. Yoshizawa}

\address{Neutron Scattering Laboratory, Institute for Solid State
Physics, University of Tokyo, Tokai, Ibaraki, 319-1106, Japan}

\author{Y. Tomioka}

\address{Joint Research Center for Atom Technology (JRCAT), Tsukuba,
Ibaraki 305-8562, Japan}

\author{Y. Tokura}

\address{Joint Research Center for Atom Technology (JRCAT), Tsukuba,
Ibaraki 305-8562, Japan}
\address{Department of Applied Physics, University of Tokyo, Bunkyo-ku,
Tokyo 113-8656, Japan}

\date{\today}

\maketitle

\begin{abstract}
 The melting process of the orbital order in
 Pr$_{0.5}$Ca$_{0.5}$MnO$_{3}$ single crystal has been studied in detail
 as a function of temperature by neutron diffraction.  It is
 demonstrated that a commensurate-incommensurate (C-IC) transition of
 the orbital ordering takes place in a bulk sample, being consistent
 with the electron diffraction studies.  The lattice structure and the
 transport properties go through drastic changes in the IC orbital
 ordering phase below the charge/orbital ordering temperature
 $T_{\rm CO/OO}$, indicating that the anomalies are intimately related to
 the partial disordering of the orbital order, unlike the consensus that
 it is related to the charge disordering process.  For the same $T$ range,
 partial disorder of the orbital ordering turns on the ferromagnetic spin
 fluctuations which were observed in a previous neutron scattering study.
\end{abstract}

}

%\section{Introduction}

Charge ordering is an ubiquitous phenomenon in transition metal oxides.
For example, a so-called ``CE-type'' charge ordering plays an essential
role in the Colossal Magnetoresistance (CMR) phenomenon which can be
observed in hole-doped perovskite manganites, such as
$R_{1-x}A_{x}$MnO$_{3}$, where $R$ denotes trivalent rare-earth ions
while $A$ denotes divalent alkaline-earth
ions. \cite{tomioka95,kuwahara95}  In the CE-type charge-ordered state
which appears near $x \sim 1/2$, Mn$^{3+}$ and Mn$^{4+}$ ions form a
checkerboard pattern with a $1:1$ ratio. As pointed out in pioneering
works on La$_{0.5}$Ca$_{0.5}$MnO$_{3}$ by Wollan and
Koehler \cite{wollan55} and Goodenough, \cite{goodenough55} this
charge-ordered state is accompanied with the ordering of the $e_g$
orbitals on Mn$^{3+}$ sites as well as the CE-type antiferromagnetic
(AFM) spin ordering, \cite{jirak85,yoshizawa95} thereby offering an ideal
platform to study an interplay among charge, orbital and spin degrees of
freedom.

Reflecting such multiple degrees of freedom, the CE-type
charge/orbital/spin ordering shows a complicated ordering/melting
process.  According to very recent x-ray scattering studies on
Pr$_{1-x}$Ca$_{x}$MnO$_{3}$ with $x \sim 1/2$, the simultaneous
charge/orbital long-range ordering is formed at $T_{\rm CO/OO}$.   Well
above $T > T_{\rm CO/OO}$, however, the charge/orbital short-range
correlation is developed, but the correlation length of the short-range
charge ordering $\xi_{\rm CO}$ is always longer than that of the orbital
ordering $\xi_{\rm OO}$, indicating that the charge ordering is a
driving force for the CE-type charge/orbital
ordering. \cite{zimmermann99,zimmermann_un} As for the spin
correlations, the CE-type AFM long-range spin order is established at
$T_{\rm N}$ which is far below $T_{\rm CO/OO}$, and the AFM spin
fluctuations in the charge/orbital-ordered phase is rapidly taken over
by the ferromagnetic fluctuations in the $T$ region for $T_{\rm N}
\lesssim T \lesssim T_{\rm CO/OO}$. \cite{kaji98}

One of the interesting but not well-elucidated aspects of the CE-type
charge/orbital ordering transition is the incommensurability of the
orbital ordering.  When the ideal CE-type charge/orbital order is
formed, zigzag arrangement of the ordered $d(3x^2-r^2)$ and
$d(3y^2-r^2)$ orbitals of Mn$^{3+}$ ions in the $ab$ plane and
associated lattice distortions will double the unit cell along the $b$
axis in the orthorhombic ($Pbnm$) lattice, and produce commensurate
superlattice reflections at $\bbox{q}_{\rm OO} = (0,\,1/2,\,0)$.
\cite{jirak85,radaelli97}  In contrast to these expectations, based on
electron diffraction studies, it has been repeatedly reported that the
CE-type charge/orbital ordering in La$_{0.5}$Ca$_{0.5}$MnO$_{3}$ and
$R_{1-x}$Ca$_{x}$MnO$_{3}$ with $x \sim 1/2$ ($R$: rare-earth ions) is
incommensurate,
\cite{radaelli97,chen96,barnabe98,mori98,chen99,mori99,jirak00} and an
elaborate charge/orbital ordering (or melting) process was suggested.
In particular, it was argued that the onset of the commensurate orbital
ordering is decoupled from the onset of the charge ordering, and the
orbital ordering can be incommensurate immediately below $T_{\rm
CO/OO}$.  In fact, the incommensurability \cite{com1} $\epsilon$ in
Pr$_{0.5}$Ca$_{0.5}$MnO$_{3}$ observed by electron diffraction becomes
finite above the AFM spin ordering temperature $T_{\rm N}$, and
$\epsilon$ grows with increasing $T$, reaching as large as $0.11 \sim
0.12$ near $T_{\rm CO/OO}$.
\cite{barnabe98,mori98,chen99,mori99,jirak00}  It should be noted that
such a large value of $\epsilon$ has been observed mainly in electron
diffraction studies or an x-ray study with La$_{0.5}$Ca$_{0.5}$MnO$_{3}$
powder sample.

By contrast, very recent x-ray scattering studies on single crystal
samples of Pr$_{1-x}$Ca$_{x}$MnO$_{3}$ argued that the orbital ordering
wave vectors remain strictly commensurate throughout the ordered phase
at any temperature, being in striking disagreement with the electron
diffraction studies. \cite{zimmermann_un,shimomura00} Incidentally,
there is no explicit statement on the incommmensurability from neutron
diffraction studies to our knowledge, seemingly giving an impression
that the charge/orbital ordering may be commensurate.  A lack of
consensus on this issue over three diffraction techniques may give rise
to a question whether the incommensurate orbital ordering is an
intrinsic bulk property.

To unravel this issue, neutron diffraction would be a key experimental
technique because of high transmissibility of neutrons which allows to
observe bulk properties of the specimen. Moreover, the neutron
diffraction has an advantage that it is sensitive to the displacements
of oxygen ions which are introduced by the orbital ordering. In what
follows, we shall report a detailed neutron diffraction study of the
melting process of the orbital ordering in a single crystal sample of
Pr$_{0.5}$Ca$_{0.5}$MnO$_{3}$. We shall demonstrate that the orbital
ordering is indeed {\it incommensurate} for certain temperature range
below $T_{\rm CO/OO}$, and argue that the anomalies are intimately
related to the disordering process of the orbital order, unlike the
conventional picture that it is related to the collapse of the charge
ordering.

Pr$_{0.5}$Ca$_{0.5}$MnO$_{3}$ is orthorhombic, but is very close to
cubic. \cite{jirak85,jirak00}  Throughout this paper, we employ the
$Pbnm$ setting for convenience of easier comparison with preceding
works.  Thereby, the lattice constants are related to the simple cubic
lattice parameter $a_c$ as $a \sim b \sim c/\sqrt{2} \sim \sqrt{2}a_c
\sim 5.4$ \AA.  In the $Pbnm$ setting, the superlattice reflections due
to the orbital ordering appear at $(h,\,k/2,\,l)$ with $k = \mbox{odd
integer}$ at low temperatures. The CE-type AFM Bragg reflections by the
Mn$^{3+}$ moments appear at $(h/2,~k,~l)$ with $k = \mbox{integer}$ and
$h,\,l=\mbox{odd integer}$, while those by the Mn$^{4+}$ moments appear
at $(h/2,\,k/2,\,l)$ with $h,\,k,\,l = \mbox{odd integer}$.

%\section{Experimental procedure}

The single crystal sample was melt-grown by the floating zone method as
described elsewhere. \cite{tomioka95} The quality of the sample was
checked by x-ray powder diffraction measurements and by inductively
coupled plasma mass spectroscopy (ICP). Neutron diffraction experiments
were performed using a triple axis spectrometer GPTAS installed at the
JRR-3M reactor in JAERI, Tokai, Japan. The incident neutron momentum is
$k_{\rm i} = 2.66$ \AA$^{-1}$ with a pyrolytic graphite (PG) filter
employed before the sample to suppress higher-order contaminations. Most
of the measurements were carried out with a double axis configuration
with 20$^{\prime}$-20$^{\prime}$-20$^{\prime}$ collimators, while for
measurements at high temperature ($T \ge 250$ K) the collimation was
relaxed to 20$^{\prime}$-40$^{\prime}$-20$^{\prime}$ or
20$^{\prime}$-40$^{\prime}$-40$^{\prime}$ to improve the visibility of
 weak superlattice peaks. The sample was mounted in an Al
can filled with He gas, and was attached to the cold head of a
closed-cycle helium gas refrigerator. The temperature was controlled
within an accuracy of 0.2 K. All the measurements were carried out on
the $(h,\,k,\,0)$ scattering plane except for the $T$ dependence of the
AFM Bragg reflection (0.5, 0.5, 1).

%\section{Results and discussion}

%\subsection{Figure 1: Temperature dependence of lineshape}

\begin{figure}[b]
 \centering \leavevmode
 \epsfxsize=85mm
 \epsfbox{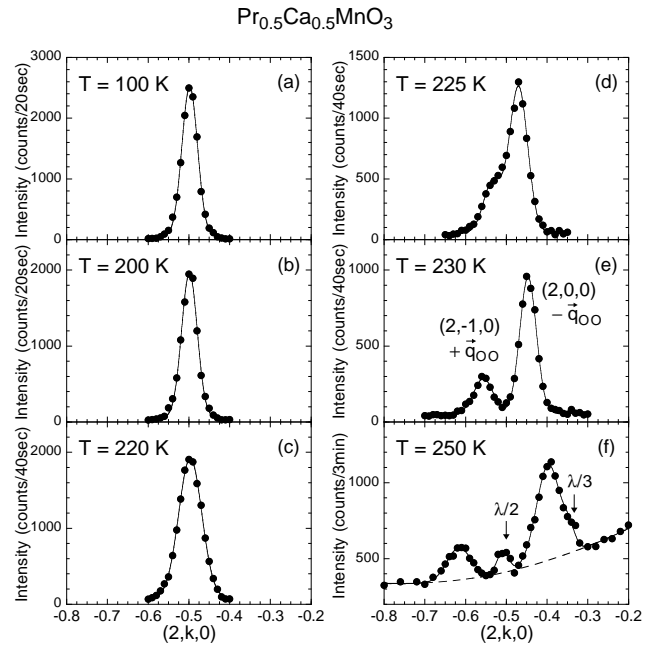}
 \vspace{2mm}
 \caption{Temperature dependence of the orbital ordering superlattice
 peak observed along the [010] direction at selected $T$'s. Note that
 $T_{\rm N} \sim 180$ K and $T_{\rm CO/OO} \sim 240$ K,
 respectively. The peak at $k=-1/2$ and the shoulder at $k=-1/3$ in (f)
 are contaminations due to the higher-order reflections.}
 \label{fig_prof}
\end{figure}

\begin{figure}[t]
 \centering \leavevmode
 \epsfxsize=85mm
 \epsfbox{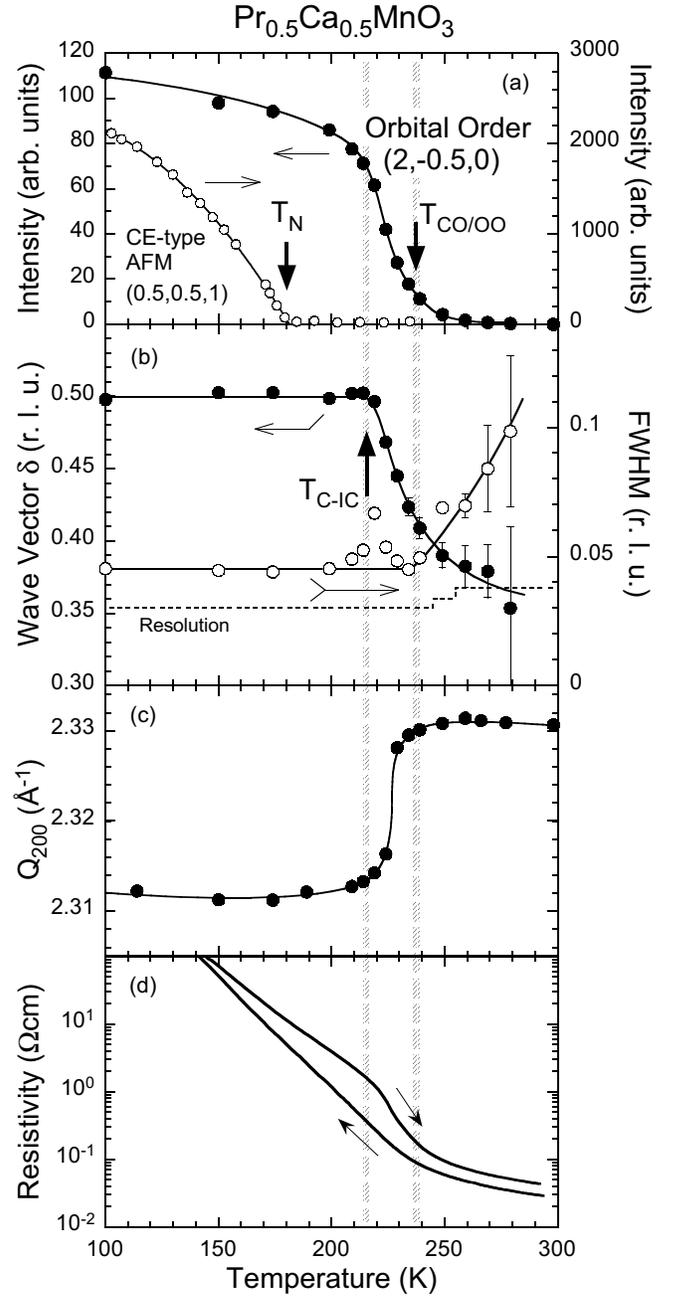}
 \vspace{2mm}
 \caption{$T$ dependences of (a): Integrated intensities for
 $(2,\,-0.5,\,0)$ orbital order peak (closed symbols) and
 $(1/2,\,1/2,\,1)$ AFM peak (open symbols). (b): Orbital order wave
 vector $\bbox{q}_{\rm OO} = (0,\,\delta,\,0)$ (closed symbols) and width
 (FWHM) of $(2,\,-0.5,\,0)$ orbital order peak (open symbols). (c):
 Scattering vector of the $(2,\,0,\,0)$ fundamental reflection. (d):
 Resistivity. In (b), the instrumental resolution is indicated as a
 dotted line. All data except the resistivity were measured with
 heating.}
 \label{fig_Tdep}
\end{figure}

In order to confirm the presence of the superlattice peaks due to
lattice modulations induced by the orbital ordering, we surveyed the
$(h,\,k,\,0)$ scattering plane at 100 K.  The superlattice peaks were
observed at the $(h,\,k \pm 1/2,\,0)$ positions with $h,\,k
=\mbox{integer}$.  The peak intensity shows a strong $Q$ dependence due
to the structure factor, and for an accessible $Q$ range of the present
experiments, the strong peaks were observed along the $(2,\,k,\,0)$
line, and the profiles around $(2,\,-0.5,\,0)$ scanned along $k$ at
selected $T$'s are shown in Fig.\ \ref{fig_prof}. Since the
charge/orbital ordering transition is weakly first order with hysteresis
(See Fig.\ \ref{fig_Tdep}(d)), all the measurements in the present study
were carried out with elevating $T$. The magnetic and charge/orbital
ordering transition temperatures for the present sample are $T_{\rm N}
\sim 180$ K and $T_{\rm CO/OO} \sim 240$ K, respectively (See Fig.\
\ref{fig_Tdep}(a)).  In the left column of Fig.\ \ref{fig_prof}, one can
see a well-defined peak centered at $k = -0.5$ up to $T \sim 220$ K,
indicating that the orbital order is commensurate.  Even for the
moderate momentum resolution of neutron diffraction, the peak has a
finite width, indicating that the orbital ordering is not a true long
range order.  From the width of profiles, the correlation length $\xi$
for the orbital ordering is roughly estimated to be an order of $\sim
100$ {\AA} at 100 K, being consistent with the x-ray scattering study
($\xi = 160 \pm 10$ {\AA}). \cite{zimmermann_un} The finite correlation
length of the orbital ordering also limits the correlation length of the
accompanied CE-type AFM spin ordering, as was demonstrated in recent
neutron diffraction studies. \cite{kaji98,radaelli97,jirak00}

At $T=220$ K, the width is slightly wider than those at lower $T$'s, but
it recovers the low $T$ value for $225~\mbox{K} \lesssim T \leq T_{\rm
CO/OO}$, and the profile splits into two peaks at the incommensurate
positions $(2,\,-0.5 \pm \epsilon,\,0)$. \cite{com1}  Upon elevating $T$
above $ T_{\rm CO/OO} \sim 240$ K, the width increases rapidly, while
the intensity decreases drastically.  As shown in Fig.\
\ref{fig_prof}(f), the intensity of the superlattice peaks at $T=250$ K
is now weaker than a tail of the Huang scattering centered at
$Q=(2,\,0,\,0)$ which was observed in a recent x-ray measurement.
\cite{shimomura00,com2}  As labeled in Fig.\ \ref{fig_prof}(e), the
weak peak at left and the stronger peak at right are assigned to
$(2,\,-1,\,0) + \bbox{q}_{\rm OO}$ and $(2,\,0,\,0) -\bbox{q}_{\rm OO}$
with $|\bbox{q}_{\rm OO}| = 0.445$, respectively.  From the scattering
configuration, it is straightforward to see that the superlattice
intensity results from the transverse component of the lattice
distortions induced by the orbital ordering because of the $|\bbox{Q}
\cdot \bbox{\eta}|^2$ term in the cross section, where $\bbox{\eta}$
represents a displacement vector of constituent ions.  Furthermore, an
asymmetry of superlattice intensities reflects the difference of the
structure factor of generic fundamental nuclear Bragg reflections, weak
$(2,\,-1,\,0)$ and intense $(2,\,0,\,0)$ reflections. It should be noted
that similar two peak profiles with asymmetric intensity in
Pr$_{0.5}$Ca$_{0.5}$MnO$_{3}$ were clearly observed in electron
diffraction studies, \cite{barnabe98,mori99} but not observed in recent
x-ray studies. \cite{zimmermann_un,shimomura00}

%\subsection{Figure 2: Summary of width, intensity, and incommesurability}

%\subsubsection{Figure 2: details of analysis}

Now that the incommesurability of the orbital ordering has been
confirmed by the present neutron diffraction study, we shall examine the
melting process of the orbital ordering in further detail.  In Fig.\
\ref{fig_Tdep}(a) are plotted the $T$ dependences of the order parameter
of the CE-type AFM spin ordering for the Mn$^{4+}$ moments observed at
$\bbox{Q}=(0.5,0.5,1)$ and that of the orbital ordering observed around
$\bbox{Q}=(2, -0.5,0)$ with elevating $T$.  The CE-type AFM spin
ordering disappears at $T_{\rm N} \sim 180$ K.  The $T$ dependence of
the orbital order parameter in Fig.\ \ref{fig_Tdep}(a) exhibits a clear
change at $T_{\rm C-IC} \sim 215$ K.  The intensity decreases gradually
up to $T_{\rm C-IC}$, then drops steeply towards $T_{\rm CO/OO} \sim
240$ K.  Interestingly, this changeover of the $T$ dependence of the
intensity corresponds to the commensurate-to-incommensurate transition
of the orbital ordering vector $\bbox{q}_{\rm OO} = (0,\,\delta,\,0)$,
as is shown in Fig.\ \ref{fig_Tdep}(b).  Up to $T_{\rm C-IC} \sim 215$
K, the peak position is commensurate with $\delta=0.5$, then $\delta$
gradually decreases towards $\delta \sim 1/3$. Namely, the orbital
ordering in Pr$_{0.5}$Ca$_{0.5}$MnO$_{3}$ is incommensurate for $T_{\rm
C-IC} \lesssim T \lesssim T_{\rm CO/OO}$. The $T$ dependence of the
width of the superlattice peak is also shown in the same panel.  The
width shows a subtle increase around $T_{\rm C-IC}$ which is also
evident in the profile shown in Fig.\ \ref{fig_prof}(c), but recovers
and maintains the low $T$ value up to $T_{\rm CO/OO}$.  The $T$ region
of the incommensurate orbital ordering corresponds exactly to the
precipitous change of the lattice constants as evidenced by the $T$
dependence of the scattering vector for the $(2,\,0,\,0)$ nuclear Bragg
reflection (Fig.\ \ref{fig_Tdep}(c)), and a steep decrease of the
resistivity (Fig.\ \ref{fig_Tdep}(d)).

%\subsection{Discussion}

%\subsubsection{Discussion: picture of the melting process}

From the data presented in Fig.\ \ref{fig_Tdep}, the following
intriguing picture emerges for the melting process of the orbital
ordering in Pr$_{0.5}$Ca$_{0.5}$MnO$_{3}$.  The melting process consists
of three stages.  First of all, the C-IC transition of the orbital
ordering is not correlated to the CE-type spin ordering at $T_{\rm N}
\sim 180$ K, and begins at $T_{\rm C-IC} \sim 215$ K.  In a very early
stage of the C-IC transition, the superlattice is still almost
commensurate, but its width increases (Fig.\ \ref{fig_prof}(c)),
indicating that the commensurate and nearly-commensurate orbital orders
are mixed in the system.  This stage may be attributed to the situation
in which occasional discommensurations are introduced to the system.
According to high resolution lattice images, it is suggested that
unpaired Mn$^{3+}$ stripes cause discommensurations.
\cite{mori98,chen99}  In the second stage of the transition, the
superlattice peak splits into two incommensurate peaks.  At the same
time, the lattice constants show precipitous changes, and the
incommensurability varies quickly, although the width recovers the low
$T$ value.  The decrease of the superlattice intensity above $T_{\rm
C-IC}$ indicates that the number of the orbital-disordered Mn$^{3+}$
ions rapidly increases in the system.  Concomitantly, the local
melting of the orbital ordering triggers the rearrangement of the local
Jahn-Teller lattice distortions, and causes the change of lattice
constants (Fig.\ \ref{fig_Tdep}(c)).  This rearrangement process also
triggers the steep decrease of the resistivity (Fig.\
\ref{fig_Tdep}(d)).  In the third stage above $T_{\rm CO/OO}$, the
charge ordering and quasi-long-range orbital ordering is destroyed, and
their correlation lengths rapidly increase.  By $T_{\rm CO/OO}$, the
local Jahn-Teller lattice distortions are sufficiently relaxed, then a
rapid change of the lattice constants disappears above $T_{\rm CO/OO}$
(Fig.\ \ref{fig_Tdep}(c)).

It should be noted that in a study of the CE-type ordering the charge
and orbital orderings are often considered to take place at the same
temperature, because the CE-type charge and orbital orders are strongly
coupled and their transition temperatures are very close. By careful
examination of the melting process of the orbital ordering, however, we
succeeded in establishing that the anomalies in the structural,
transport and magnetic properties take place in the IC orbital ordering
state below the charge ordering temperature $T \sim T_{\rm CO/OO}$. The
present results indicate that the importance of distinguishing the
gradual melting process of orbital ordering from the collapse of charge
ordering in order to understand the nature of the physical properties in
the CMR manganites.

%\subsubsection{Discussion: relation of $T_{\rm N}$ and the onset of
%I-IC, and influence of magnetic fluctuations}

It is also noteworthy that the partial disordering of the orbital
order is consistent with the onset of ferromagnetic spin fluctuations
previously reported. \cite{kaji98}  Kajimoto {\it et al.} reported that
the ferromagnetic spin fluctuations are developed for $T_{\rm N}
\lesssim T \lesssim T_{\rm CO/OO}$.  In the melting process of the C-IC
orbital ordering, a partially disordered orbital order disturbs static
AFM exchange paths, and gives rise to ferromagnetic spin fluctuations.
In this sense, the C-IC orbital disordering process is intimately
related to the ferromagnetic spin fluctuations.

In summary, we have studied the melting process of the orbital ordering
in Pr$_{0.5}$Ca$_{0.5}$MnO$_{3}$ single crystal as a function of $T$ by
neutron diffraction measurements.  We successfully demonstrated that an
incommensurate orbital ordering certainly exists in a bulk sample.  The
quasi long-range orbital ordering persists well above the C-IC
transition temperature $T_{\rm C-IC}$. The incommensurate melting of the
orbital order induce the rearrangement of the lattice structure and the
steep decrease of the resistivity due to partial disordering of the
orbital order, and also induce ferromagnetic spin fluctuations which was
reported in the previous neutron scattering study. This fact indicates
that the drastic anomalies in charge/orbital ordered manganites are
caused by the change of the orbital order rather than the charge order.

%\section*{Acknowledgement}

This work was supported by a Grand-In-Aid for Scientific Research from
the Ministry of Education, Science, Sports and Culture, Japan and by the
New Energy and Industrial Technology Development Organization (NEDO) of
Japan.

% References


\begin{references}
\bibitem{com1}  The incommensurability $\epsilon$ is defined as
 $\bbox{q}_{\rm OO} = (0,\,1/2 - \epsilon,\,0)$ from the observed
 position of the superlattice peaks.
\bibitem{com2}  The scan profile consists of two components: one is the
 superlattice peak of the orbital ordering, and the other is the Huang
 scattering.  Since they have completely different physical origins,
 they should be strictly distinguished.
\end{references}
\end{document}